\documentclass{elsart}
\journal{New Astronomy}

\usepackage{natbib}
\usepackage{psfig}
\usepackage{amssymb}

\begin{document}

%%%%%%%%%%%%%%%%%%%%%%%%%%%%%%%%%%%%%%%%%%%%%%%%%%%%%%%%%%%%%%%%%%%%%%%%

% User defined commands/abbreviations/symbols:
% Short-hand notations
\newcommand{\n}		{\noindent}%
\newcommand{\sn}	{\smallskip\noindent}%
\newcommand{\ve}	{\vfill\eject}%
\renewcommand{\etal}	{\mbox{et al.\,}}%
% Symbols
\newcommand{\cge}	{\ensuremath{\,\gtrsim\,}}%
\newcommand{\cle}	{\ensuremath{\,\lesssim\,}}%
\newcommand{\bul}	{\ensuremath{\bullet}\ }%
\newcommand{\Msun}	{\ensuremath{M_{\odot}}}%
\newcommand{\zform}	{\ensuremath{\mbox{\rm z}_{\mbox{\rm\scriptsize form}}}}%
\newcommand{\arcmpt}	{{$\buildrel{\prime}      \over .$}}
\newcommand{\arcspt}	{{$\buildrel{\prime\prime}\over .$}}
\newcommand{\re}	{\ensuremath{r_e}}%
\newcommand{\micron}	{\ensuremath{\mu}\mbox{\rm m}}%

%%%%%%%%%%%%%%%%%%%%%%%%%%%%%%%%%%%%%%%%%%%%%%%%%%%%%%%%%%%%%%%%%%%%%%%%%%%%

\begin{frontmatter}

\title {How JWST can measure First Light,\\ 
	Reionization and Galaxy Assembly}

\author{Rogier A. Windhorst, Seth H. Cohen, Rolf A. Jansen}
\address{Department of Physics and Astronomy, Arizona State University,
	Box 871504, Tempe, AZ 85287;\ \ Email:\ Rogier.Windhorst@asu.edu}
%\ead[url]{http://www.asu.edu/clas/hst/www/jwst/}

\author {Chris Conselice \& Haojing Yan}
\address {California Institute of Technology, MS 100-22, Pasadena, CA 91125}

\begin{abstract}
We summarize the design and performance of the James Webb Space Telescope
that is to be launched to an L2 orbit in 2011, and how it is designed, in
particular, to study the epochs of First Light, Reionization and Galaxy
Assembly. 
\end{abstract}


\end{frontmatter}

%%%%%%%%%%%%%%%%%%%%%%%%%%%%%%%%%%%%%%%%%%%%%%%%%%%%%%%%%%%%%%%%%%%%%%%%

\section{The James Webb Space Telescope and its Instruments }

\n The James Webb Space Telescope (JWST) is currently designed as a
fully deployable 6.5~meter segmented IR telescope, optimized for imaging
and spectroscopy from 0.6~\micron\ to 28~\micron, to be launched by NASA
in 2011 (Mather \& Stockman 2000).  After its launch --- currently
planned with an Ariane V --- JWST will make a several month journey to
the Earth--Sun Lagrange point L2.  En route, JWST will be automatically
deployed in phases, its instruments will be tested, and will be inserted
into an L2 halo orbit.  JWST has a nested array of sun-shields to keep
its ambient temperature at \cle 40~K, allowing faint imaging (to AB\cle
31.5~mag or $\simeq$1~nJy) and spectroscopy (AB\cle 29~mag) in the near-
to mid-IR.  From L2, JWST can cover the whole sky in segments that move
along in RA with the Earth.  It will have an observing efficiency \cge 
70\%, and send data back to Earth every day. 

The Optical Telescope Element (OTE) of the 6.5~meter JWST has 18
hexagonal mirror segments --- each of 1.3~m diameter --- and a total
edge-to-edge diameter of 6.60~m.  Its effective circular diameter is
5.85~m, and its effective collecting area is 25~m$^2$.  Instead of the
original design of 36 smaller segments, the OTE design converged on 18
larger segments.  This had both risk- and cost-benefits, was easier to
construct, and provided smaller mid-spatial frequency OTE errors,
resulting in better 1.0~\micron\ performance.  With only 18 mirror
segments, one cannot cleanly descope the telescope aperture without
doing major harm to its resulting PSF.  Fig.~1a shows a simulation of
the 6.5~m JWST PSF, and Fig.~1b shows a 20~hr JWST simulation that
reaches the HUDF depth.  The JWST science requirements and its
instruments are described by Gardner \etal (2004), M.~Rieke (2005, this
Vol.), and on the websites listed in the References.  In summary, JWST
will have the following instruments:

\bul {\bf NIRCam:}\ The Near-Infrared Camera is made by an UofA +
Lockheed + CSA consortium.  NIRCam will do imaging from
0.6--5.3~\micron\ using a suite of broad-, medium-, and narrow-band
filters.  It uses two identical and independently operated imaging
modules, with two wavelengths observable simultaneously via a dichroic
that splits the beam around 2.35~\micron.  Each of these two channels
has an independently operated 2\arcmpt 2$\times$4\arcmpt 6 field-of-view
(FOV).  Both channels are Nyquist-sampled: the short wavelength channel
at 2.0~\micron\ with 0\arcspt 0317/pixel, and the long wavelength at
4.0~\micron\ with 0\arcspt 0648/pixel.  NIRCam's ten 2k$\times$2k
HgCdTe arrays will be passively cooled.  As shown below, a large FOV is
essential to detect the expected First Light objects. 

\bul {\bf NIRSpec:}\ The Near-Infrared Spectrograph is made by an ESA +
GSFC consortium.  NIRSpec will do spectroscopy with resolving powers of
R$\sim$100 in prism mode, of R$\sim$1000 in multi-object mode using a
micro-electromechanical array system (MEMS) of micro-shutters that can
open slitlets on previously imaged known objects, and of R$\sim$3000
using long-slit spectroscopy.  All NIRSpec spectroscopic modes have a
$\sim$3\arcmpt 4$\times$3\arcmpt 4 FOV.  NIRSpec also has an IFU. 

\bul {\bf MIRI:}\ The Mid-InfraRed Instrument is made by a UofA + JPL +
ESA consortium.  MIRI will do imaging and spectroscopy from
5--28~\micron.  It is passively cooled by a cryocooler instead of a
cryostat, which would have only a 5 year lifetime.  The NIRCam and MIRI
sensitivity complement each other, straddling 5~\micron\ in wavelength. 
To detect {\it and} confirm the first star-forming objects at redshifts
z$\,\simeq\,$15--20 in \cge 10$^5$~sec (28~hrs) integration times, JWST
needs both NIRCam at 0.6--5~\micron\ and MIRI at 5--28~\micron. 

\bul {\bf FGS:}\ The Fine Guidance Sensor is made by CSA and provides
stable pointing at the milli-arcsecond level.  It will have sufficient
sensitivity and a large enough FOV to find guide stars with \cge 95\%
probability at any point in the sky.  The FGS will have three
simultaneously imaged fields of view of 2\arcmpt 3$\times$2\arcmpt 3,
one of which feeds a pure guider channel, one feeds a guider channel
plus a long-wavelength R$\sim$100 tunable-filter channel with light
split by a dichroic, and another feeds a short-wavelength R$\sim$100
tunable-filter channel. 

\ve 

%%%%%%%%%%%%%%%%%%%%%%%%%%%%%%%%%%%%%%%%%%%%%%%%%%%%%%%%%%%%%%%%%%%%%%%%
%   FIGURE 1   FIGURE 1   FIGURE 1   FIGURE 1   FIGURE 1   FIGURE 1    %
%%%%%%%%%%%%%%%%%%%%%%%%%%%%%%%%%%%%%%%%%%%%%%%%%%%%%%%%%%%%%%%%%%%%%%%%
\n\makebox[\textwidth]{
   \psfig{file=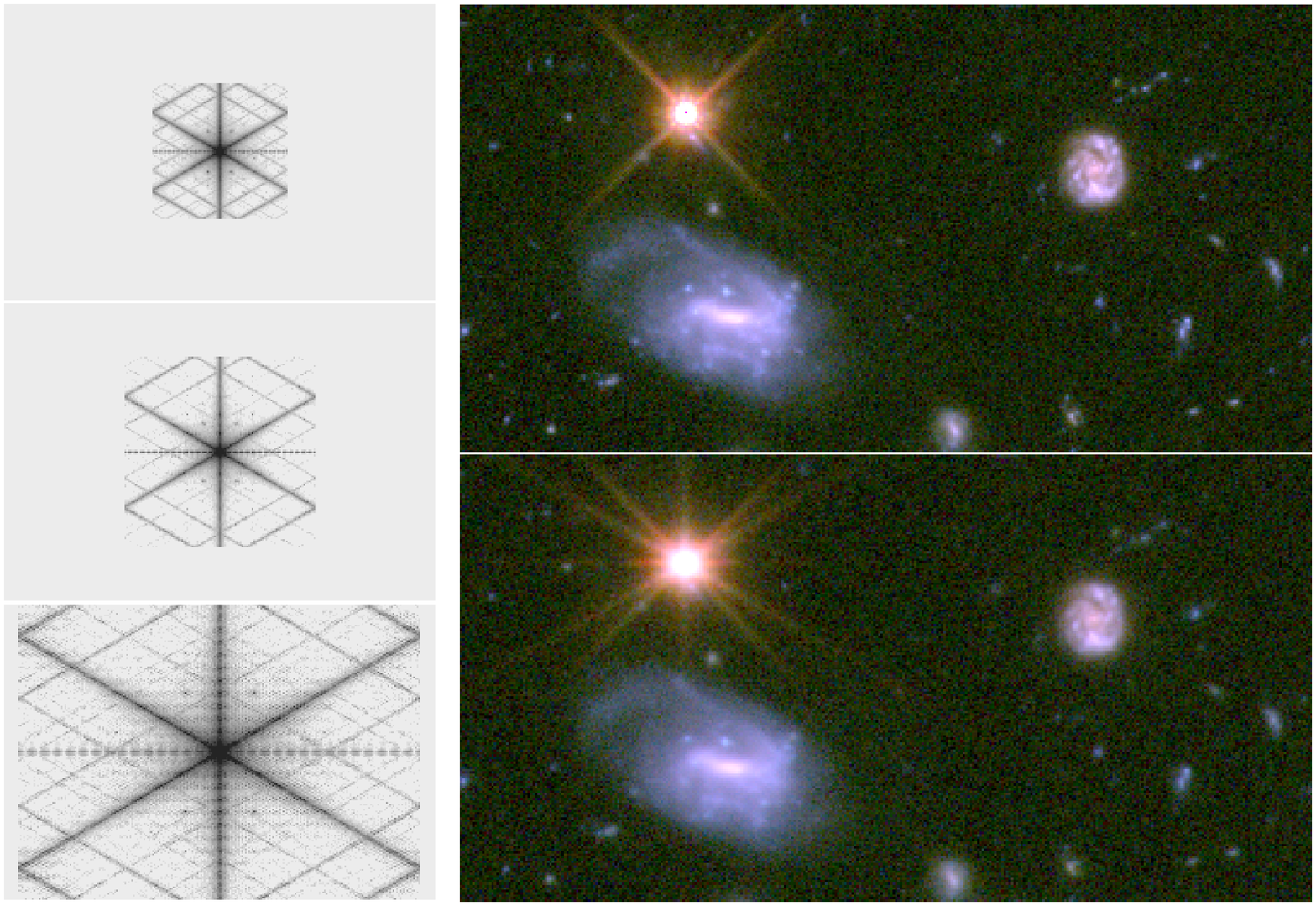,width=\textwidth,angle=0}\ \ 
}

\n {\footnotesize\baselineskip=10pt
{\bf Fig.~1a {\sc (left)}.}\ JWST PSF's at 0.7~\micron\ (\emph{top}), 
1.0~\micron\ (\emph{middle}), and 2.0~\micron\ (\emph{bottom}), simulated
by Ball Aerospace and GSFC for an OTE wave-front error WFE$<$150 nm, 
corresponding to an encircled energy of EE=74\% inside a $r$=0\arcspt 15
radius at 1.0~\micron.} 
\n {\footnotesize\baselineskip=10pt 
{\bf Fig.~1b {\sc (right)}.}\ 240~hrs HST/ACS Vi'z' in the HUDF
(\emph{top}), and a JWST/NIRCam simulation at 0.7, 1.0, 2.0~\micron,
20~hrs in total (\emph{bottom}). The ACS PSF is ignored, resulting in
three extra diffraction spikes for the bright star.  The color image can
be viewed at very high resolution from the PDF file on our 
website$^{[31]}$. Even with a JWST PSF meeting current specs at
0.7 \& 1.0~\micron\ (74\% EE inside $r$=0\arcspt 15 at 1.0~\micron), some
loss in contrast is seen for faint ``blue'' objects.  This is further
quantified by Windhorst \etal (2003, 2005).}
%%%%%%%%%%%%%%%%%%%%%%%%%%%%%%%%%%%%%%%%%%%%%%%%%%%%%%%%%%%%%%%%%%%%%%%%

\n JWST has fully redundant imaging and spectroscopic modes.  It will
not be serviced at L2, and will undergo an extensive series of
ground-testing and thermal vacuum testing in 2008--2009, after its main
design and construction phase in 2004--2008.  The main NASA contractor
is Northrop Grumman Space Technology (``NGST'') in Redondo Beach (CA).

\section{Measuring First Light, Reionization \& Galaxy Assembly}

\n {\bf \bul First Light:}\ The WMAP polarization results suggested that
the universe was first reionized at redshifts as early as
z$\,\simeq\,$20 (Spergel \etal 2003).  This epoch of First Light is
thought to have started with Population~III stars of\linebreak

\ve 

%%%%%%%%%%%%%%%%%%%%%%%%%%%%%%%%%%%%%%%%%%%%%%%%%%%%%%%%%%%%%%%%%%%%%%%%
%   FIGURE 2   FIGURE 2   FIGURE 2   FIGURE 2   FIGURE 2   FIGURE 2    %
%%%%%%%%%%%%%%%%%%%%%%%%%%%%%%%%%%%%%%%%%%%%%%%%%%%%%%%%%%%%%%%%%%%%%%%%
\n\makebox[\textwidth]{
   \psfig{file=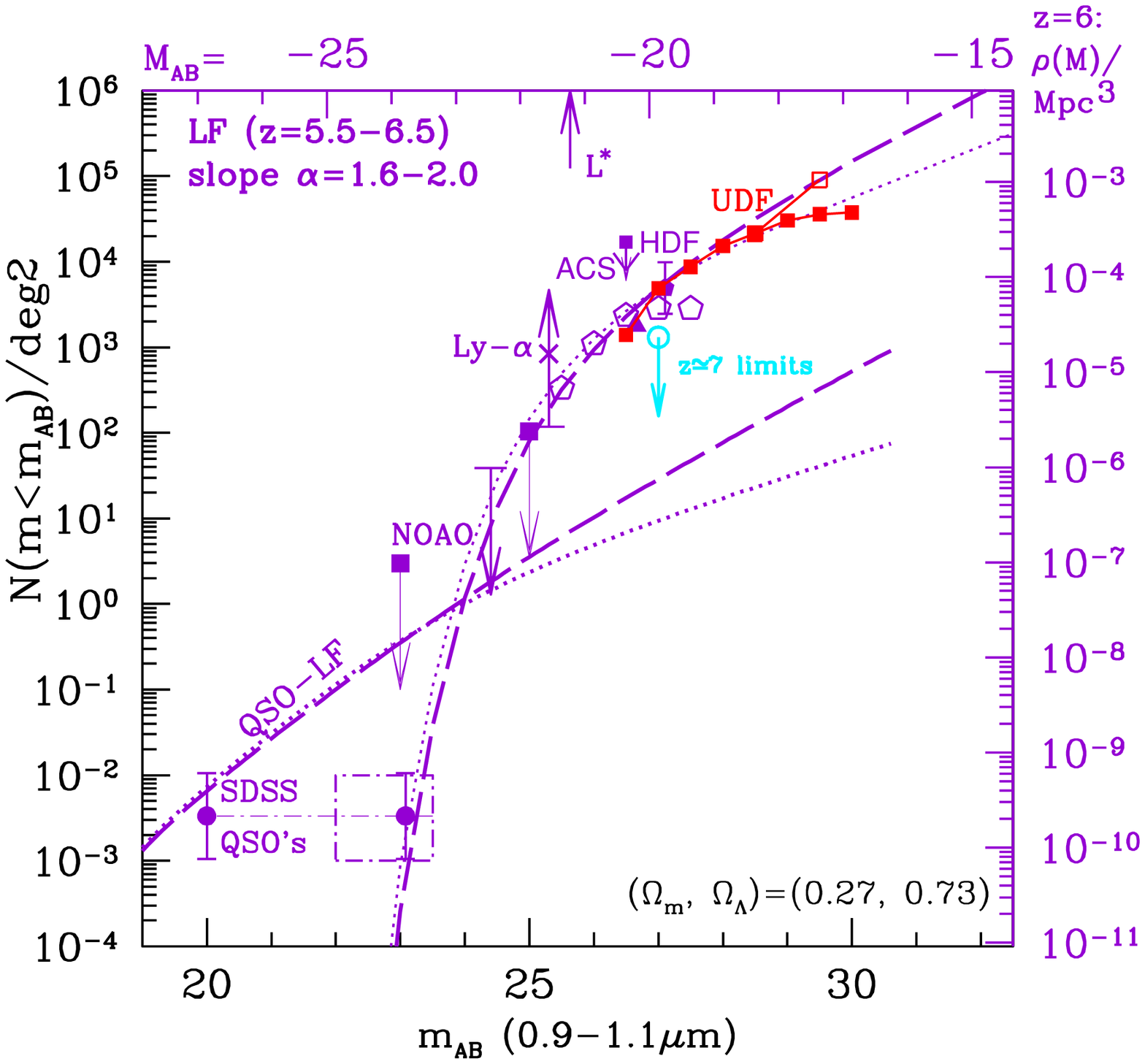,width=0.305\textheight}
   \psfig{file=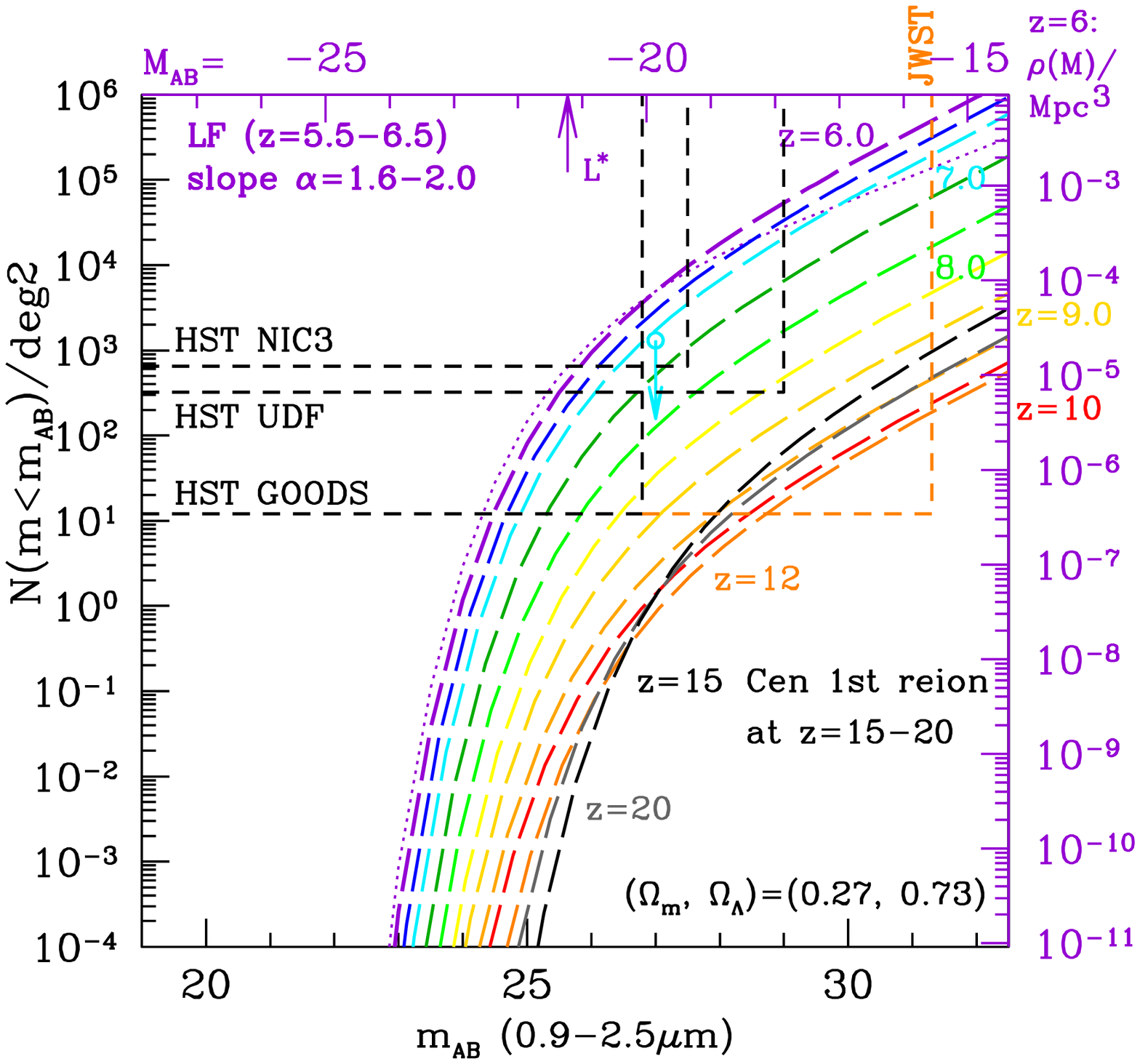,width=0.320\textheight}
}
\n {\footnotesize\baselineskip=10pt
{\bf Fig.~2a {\sc (left)}.}\ The HUDF suggested that the luminosity
function (LF) of z$\,\simeq\,$6 objects may be very steep, with faint-end
Schechter slope $\vert$$\alpha$$\vert\simeq1.8$--1.9 (Yan \& Windhorst
2004b).  Dwarf galaxies and not quasars therefore likely completed the
reionization epoch at z$\,\simeq\,$6 (Yan \etal 2004a).  This is what JWST
likely will observe in detail and to AB=31.5~mag (1~nJy).}
\n {\footnotesize\baselineskip=10pt
{\bf Fig.~2b {\sc (right)}.}\ HST/ACS can detect objects at z\cle 6.5,
but its discovery space A$\cdot\Omega\!\cdot\!\Delta\log(\lambda)$ cannot
trace the entire reionization epoch.  NICMOS similarly is limited to 
z\cle 8--10.  JWST can trace the entire reionization epoch from First
Light at z$\,\simeq\,$20 to the end of reionization at z$\,\simeq\,$6.}
%%%%%%%%%%%%%%%%%%%%%%%%%%%%%%%%%%%%%%%%%%%%%%%%%%%%%%%%%%%%%%%%%%%%%%%%

\vspace*{-2pt}
\n 200-300\Msun\ at z$\,\simeq\,$15--25.  These Pop~III star clusters
and their extremely luminous supernovae should be visible to JWST at
z$\,\simeq\,$15--25.  This epoch may have been followed by a delayed
epoch of Pop~II star-formation, since the Pop~III supernovae may have
sufficiently heated the IGM that it could not cool and form normal
Pop~II halo stars until z$\,\simeq\,$10 (Cen 2003).  These halo Pop~II
stars may have formed in dwarf galaxies of mass $\simeq\,$$10^6$ to
$10^9\Msun$ with a gradual onset between z$\simeq$10--6.  The
reionization history may have been more complex and/or heterogeneous,
with some Pop~II stars forming in sites of sufficient density
immediately following their Pop~III predecessors.  JWST is designed to
detect the first super star-clusters and star-forming dwarf galaxies
throughout the entire reionization epoch from z$\,\simeq\,$20 to
z$\,\simeq\,$6, and measure their Luminosity Function (LF), as
illustrated in Fig.~2b, which is based on the double reionization model
of Cen (2003).  The surface densities at z$\sim$20 were independently
predicted by Stiavelli (2005, priv.\ comm.), based on the models of
Stiavelli \etal (2004).  One must keep in mind that all these
predictions are uncertain by at least 0.5~dex.  Since in WMAP cosmology
the amount of available volume per unit redshift decreases for z\cge 2,
the observed surface density of objects at z$\,\simeq\,$10--20 will be
small, as shown in Fig.~2b, depending somewhat on the exact hierarchical
model predictions used.  Hence, to observe the LF of First Light
star-clusters and subsequent dwarf-galaxy formation requires JWST to
survey GOODS-sized areas to AB=31.5 mag ($\simeq\,$1~nJy at
10-$\sigma$), using 7 filters for reliable photometric redshifts, since
objects with AB\cge 29\linebreak

\n mag will be too faint for spectroscopy.  With the 6.5~m JWST at its
current specifications, such a survey requires 0.4 years of exposure
time.  I.e., even with the current 6.5~m JWST, this goes beyond an HST
Treasury-sized program, and may require a dedicated multi-cycle
community wide effort.

\sn \bul {\bf Reionization:}\ The HUDF data suggested that the LF of
z$\,\simeq\,$6 objects is potentially very steep (Bouwens \etal 2004,
Yan \& Windhorst 2004b), with a faint-end Schechter slope
$\vert$$\alpha$$\vert\simeq1.8$--1.9 after correcting for incompleteness
of the ACS i-band dropout samples (Fig.~2a).  Deep HST/ACS grism spectra
confirmed 85--93\% of the HUDF i-band dropouts to AB=27 mag to be at
z$\,\simeq\,$6 (Malhotra \etal 2005).  The steep faint-end slope of the
z$\,\simeq\,$6 LF implies that dwarf galaxies may have collectively
provided enough UV-photons to complete reionization at z$\,\simeq\,$6
(Yan \& Windhorst 2004a).  This assumes that the Lyman continuum escape
fraction at z$\,\simeq\,$6 is as large as observed in Lyman Break
Galaxies at z$\,\simeq\,$3, which is reasonable --- although not proven
--- given the expected low dust content in dwarf galaxies at
z$\,\simeq\,$6.  Hence, dwarf galaxies, and not quasars, likely
completed the reionization epoch at z$\,\simeq\,$6.  The Pop~II stars in
dwarf galaxies cannot have started shining {\it pervasively} much before
z$\,\simeq\,$7--8, or no neutral H\,{\sc i} would be seen in the
foreground of z\cge 6 quasars (Fan \etal 2003), and so dwarf galaxies
may have ramped up their formation fairly quickly from z$\,\simeq\,$10
to z$\,\simeq\,$6.  A first glimpse of this may already be visible in
the NIC3 surveys of the HUDF, which suggests a lower surface density of
z\cge 7 candidates compared to z$\,\simeq\,$6 objects (Bouwens \etal
2004b; Yan \etal 2004b; light blue upper limit in Fig.~2a--2b), although
the \cge 600 HST orbits spent on the HUDF only resulted in a few z\cge 7
candidates at best. 

HST/ACS can detect objects at z\cle 6.5, but its discovery space
A$\cdot\Omega\!\cdot\!\Delta\log(\lambda)$ cannot trace the entire
reionization epoch.  HST/NICMOS similarly is limited to z\cle 8--10 and
has very limited statistics.  HST/WFC3 --- if it gets launched --- will
be able to explore the same redshift space, but with a wider FOV than
NICMOS/NIC3.  Fig.~2 shows that with proper survey strategy (area
\emph{and} depth), JWST can trace the entire reionization epoch from
First Light at z$\,\simeq\,$20 to the end of the reionization epoch at
z$\,\simeq\,$6.  Hence, JWST can trace the entire reionization epoch, by
detecting the first star-forming objects, and measure their LF and its
evolution.  For this to be successful in realistic or conservative model
scenarios, JWST needs to have the quoted sensitivity/aperture (``A''; to
reach AB\cge 31~mag), field-of-view (FOV=$\Omega$; to cover GOODS-sized
areas), and wavelength range (0.7--28~\micron; to cover SED's from the
Lyman to Balmer breaks at z\cge 6--20), as summarized in Fig.~2b. 

Fig.~3a shows the sum of 49 compact isolated i-band dropouts in the 
HUDF\linebreak

%%%%%%%%%%%%%%%%%%%%%%%%%%%%%%%%%%%%%%%%%%%%%%%%%%%%%%%%%%%%%%%%%%%%%%%%
%   FIGURE 3   FIGURE 3   FIGURE 3   FIGURE 3   FIGURE 3   FIGURE 3    %
%%%%%%%%%%%%%%%%%%%%%%%%%%%%%%%%%%%%%%%%%%%%%%%%%%%%%%%%%%%%%%%%%%%%%%%%
\n\makebox[\textwidth]{
   \psfig{file=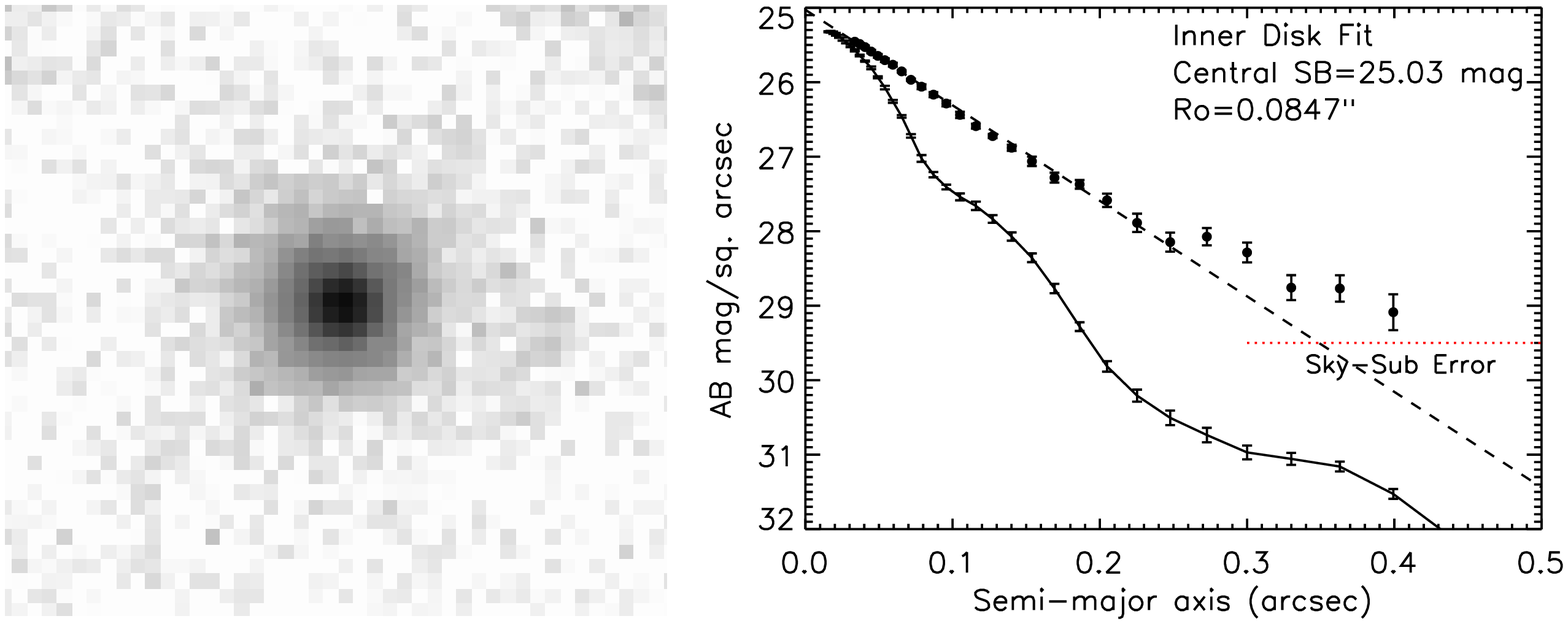,width=0.90\textwidth,angle=0}
}

\n {\footnotesize\baselineskip=10pt
{\bf Fig.~3a {\sc (left)}.}\ Sum of 49 compact isolated i-band dropouts
in the HUDF (Yan \& Windhorst 2004b).  This image is representative of a
5000~hr HST $z$-band exposure --- equivalent to a 330~hr JWST 1~\micron\
exposure --- of an average compact isolated z$\,\simeq\,$6 object.}
\n {\footnotesize\baselineskip=10pt
{\bf Fig.~3b {\sc (right)}.}\ The radial surface brightness (SB) profile
of the image stack of Fig.~3a.  The physical radius where the profile
starts to deviate from a pure exponential profile can be used as a
dynamical clock (van~Albada 1982) to constrain their dynamical age,
which is $\simeq\,$100-200~Myr at z$\,\simeq\,$6, i.e., similar to their
SED age.  HST cannot accurately measure individual objects at
z$\,\simeq\,$6, but in long integrations JWST can measure the
reionizing objects from z$\,\simeq\,$20 to z$\,\simeq\,$6.}
%%%%%%%%%%%%%%%%%%%%%%%%%%%%%%%%%%%%%%%%%%%%%%%%%%%%%%%%%%%%%%%%%%%%%%%%

\n (Yan \& Windhorst 2004b), which is a stack of about half the
z$\,\simeq\,$6 objects that have no obvious interactions or neighbors. 
These objects all have similar fluxes and half-light radii (\re), so
this image represents a 5000~hr HST/ACS z-band exposure on an ``average
compact isolated z$\,\simeq\,$6 object'', which is equivalent to a
$\sim\,$330~hr JWST 1~\micron\ exposure on \emph{one} such object. 
Fig.~3b shows that the radial SB-profile of this stacked image deviates
from a pure exponential profile for $r$\cge 0\arcspt 25, at SB-levels
that are well above those corresponding to the 3-$\sigma$ limits due to
PSF and sky-subtraction errors.  In hierarchical models, this physical
scale-length is a direct dynamical clock (e.g., van~Albada 1982),
constraining the dynamical age of these compact isolated i-band
z$\,\simeq\,$6 dropouts to $\simeq\,$100-200~Myr, i.e., similar to their
stellar population age. 

This then suggests that the bulk of their stars observed at
z$\,\simeq\,$6 may have started forming around
\zform$\,\simeq\,$7.0$\pm$0.5.  In the light of WMAP results, this is
consistent with the double reionization model of Cen (2003), where the
first reionization by Pop~III stars at z$\,\simeq\,$10--20 is followed
by a delayed onset of Pop~II star-formation in dwarf galaxies at z\cle
10.  HST cannot accurately measure individual light-profiles at
z$\,\simeq\,$6, but in long integrations JWST can measure the growth of
objects from the onset of reionization at z$\,\simeq\,$20 to its end at
z$\,\simeq\,$6.

\sn \bul {\bf Galaxy Assembly:}\ One of the remarkable discoveries of
HST was how numerous and small faint galaxies are (Abraham \etal 1995,
Glazebrook 1995).  They are likely the building blocks of the giant
galaxies seen today.  Galaxies\linebreak

%%%%%%%%%%%%%%%%%%%%%%%%%%%%%%%%%%%%%%%%%%%%%%%%%%%%%%%%%%%%%%%%%%%%%%%%
%   FIGURE 4   FIGURE 4   FIGURE 4   FIGURE 4   FIGURE 4   FIGURE 4    %
%%%%%%%%%%%%%%%%%%%%%%%%%%%%%%%%%%%%%%%%%%%%%%%%%%%%%%%%%%%%%%%%%%%%%%%%
\n\makebox[\textwidth]{
   \psfig{file=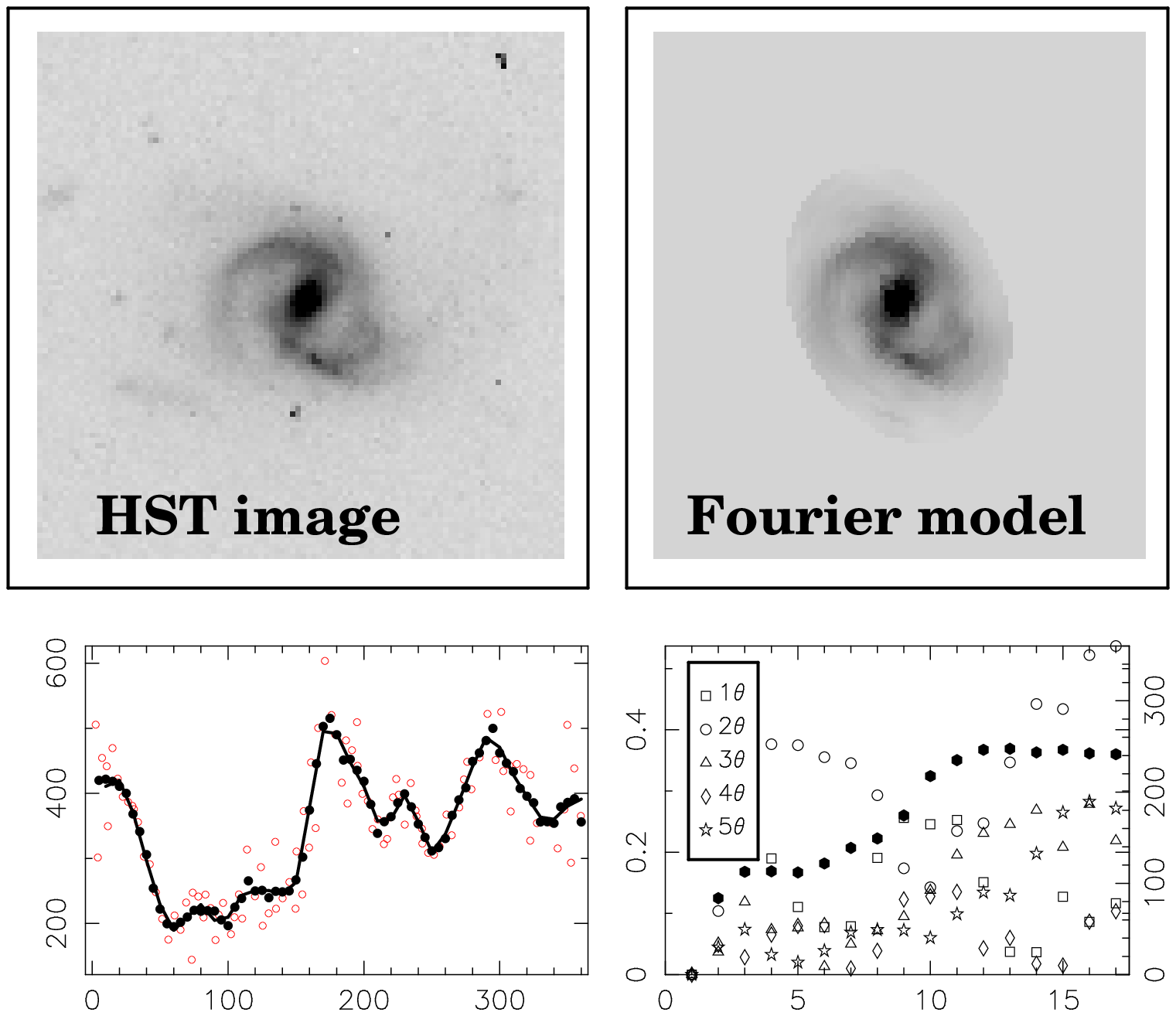,width=0.70\textwidth}
}

\n {\footnotesize\baselineskip=10pt
{\bf Fig.~4.}\ Fourier Decomposition (FD) is a robust way to measure
galaxy structure and morphology in a quantitative way (e.g., Odewahn
\etal 2002).  FD decomposes a real image (\emph{top left}) into Fourier
series in successive concentric annuli (\emph{bottom left}), resulting
in an accurate decomposition (\emph{top right}) that turns $\sim$100$^2$
pixels (\emph{top left}) into less than a 100 numbers (\emph{bottom 
right}).  The even Fourier components describe symmetric galaxy features
(arms, rings), while the odd Fourier components indicate asymmetric 
structures (lopsidedness, star forming regions, etc).  Its large 
wavelength range allows JWST to measure the evolution of each type of
physical feature directly.}
%%%%%%%%%%%%%%%%%%%%%%%%%%%%%%%%%%%%%%%%%%%%%%%%%%%%%%%%%%%%%%%%%%%%%%%%

\n with types on the present-day Hubble sequence formed over a wide
range of cosmic time, but with a notable phase transition around 
z$\,\simeq\,$1: (1) Subgalactic units rapidly merge from z$\,\simeq\,$7
to z$\,\simeq\,$1 to grow bigger units; (2) Merger products start to
settle as galaxies with giant bulges or large disks around 
z$\,\simeq\,$1.  These evolved mostly passively since then (as tempered
by the cosmological constant, see e.g., Cohen \etal 2003), resulting in
the giant galaxies that we see today.  JWST can measure how galaxies of
all types formed over a wide range of cosmic time, by accurately
measuring their distribution over rest-frame type and structure as a
function of redshift or cosmic epoch (see Fig.~4 and 5). 

Fig.~4 illustrates how Fourier Decomposition of nearby galaxies seen
with HST in the rest-frame UV can be used to quantitatively measure
bars, rings, spiral arms, and other structural features (e.g., Odewahn
\etal 2002).  Fourier Decomposition is remarkably good in distinguishing
and quantifying bars and (1-armed, 2-armed) spiral structure.  Observing
distant galaxies in deep JWST NIRCam images will in a similar fashion
directly trace the evolution of bars,

\ve 

%%%%%%%%%%%%%%%%%%%%%%%%%%%%%%%%%%%%%%%%%%%%%%%%%%%%%%%%%%%%%%%%%%%%%%%%
%   FIGURE 5   FIGURE 5   FIGURE 5   FIGURE 5   FIGURE 5   FIGURE 5    %
%%%%%%%%%%%%%%%%%%%%%%%%%%%%%%%%%%%%%%%%%%%%%%%%%%%%%%%%%%%%%%%%%%%%%%%%
\n\makebox[\textwidth]{
   \psfig{file=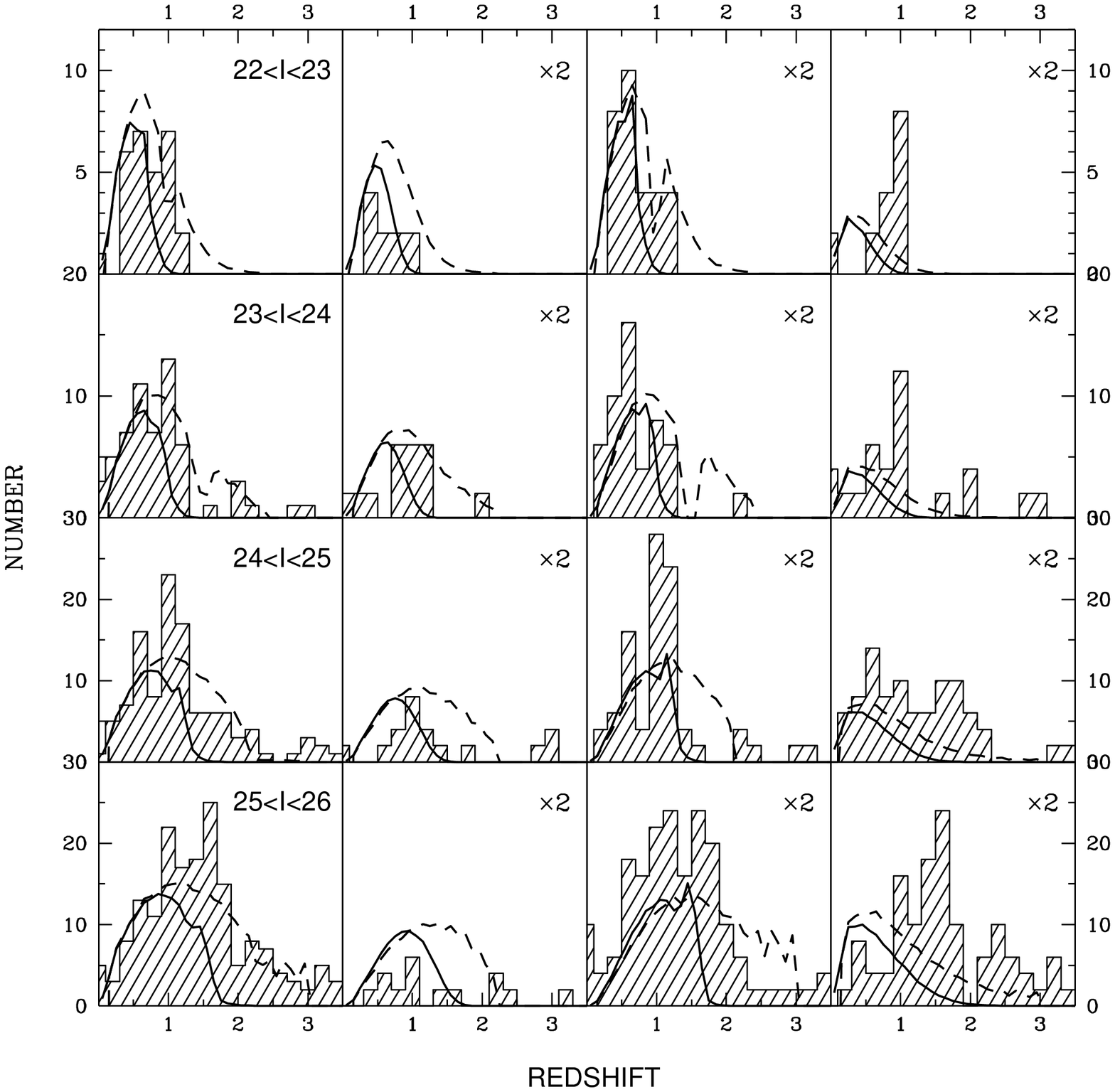,width=0.80\textwidth}
}

\n {\footnotesize\baselineskip=10pt
{\bf Fig.~5.}\ JWST can measure how galaxies of all Hubble types formed
over a wide range of cosmic time, by measuring their redshift
distribution as a function of rest-frame type (e.g., Driver \etal 1998). 
These are shown here for all objects (black histograms in left column),
early-types (E/S0; red histograms), mid-type spirals (Sa--Sc; green
histograms), and late-types (Scd--Im/Pec; blue histograms).  JWST 
``types'' will not only include galaxy morphology or Hubble class --- 
which is poorly determined at z\cge 1 --- but also physical structures,
such as disks, spiral arms, bars, asymmetry, clumpiness, etc., as 
determined by Fourier Decomposition (Fig.~4).}
%%%%%%%%%%%%%%%%%%%%%%%%%%%%%%%%%%%%%%%%%%%%%%%%%%%%%%%%%%%%%%%%%%%%%%%%

\n rings, spiral arms, and other structural features in these objects,
which will be anchored in the rest-frame UV images of the same
structures seen in nearby galaxies (Windhorst \etal 2002).  This will
allow JWST to measure the detailed history of galaxy assembly in the
epoch z$\,\simeq\,$1--3, when most of today's giant galaxies were made.

The uncertain rest-frame UV-morphology of galaxies is dominated by young
and hot stars, with often copious amounts of dust superimposed.  This
will complicate the study of NIRCam images of very high redshift 
galaxies, but the longer wavelength images from MIRI will help constrain
the effects from dust.  With good images a quantitative analysis of the
rest-frame wavelength dependent morphology and structure can be then made
(e.g., Odewahn \etal 2002).  JWST can measure how galaxies of all Hubble
types formed over a wide range

\ve 

%%%%%%%%%%%%%%%%%%%%%%%%%%%%%%%%%%%%%%%%%%%%%%%%%%%%%%%%%%%%%%%%%%%%%%%%
%   FIGURE 6   FIGURE 6   FIGURE 6   FIGURE 6   FIGURE 6   FIGURE 6    %
%%%%%%%%%%%%%%%%%%%%%%%%%%%%%%%%%%%%%%%%%%%%%%%%%%%%%%%%%%%%%%%%%%%%%%%%
\n\makebox[\textwidth]{
   \psfig{file=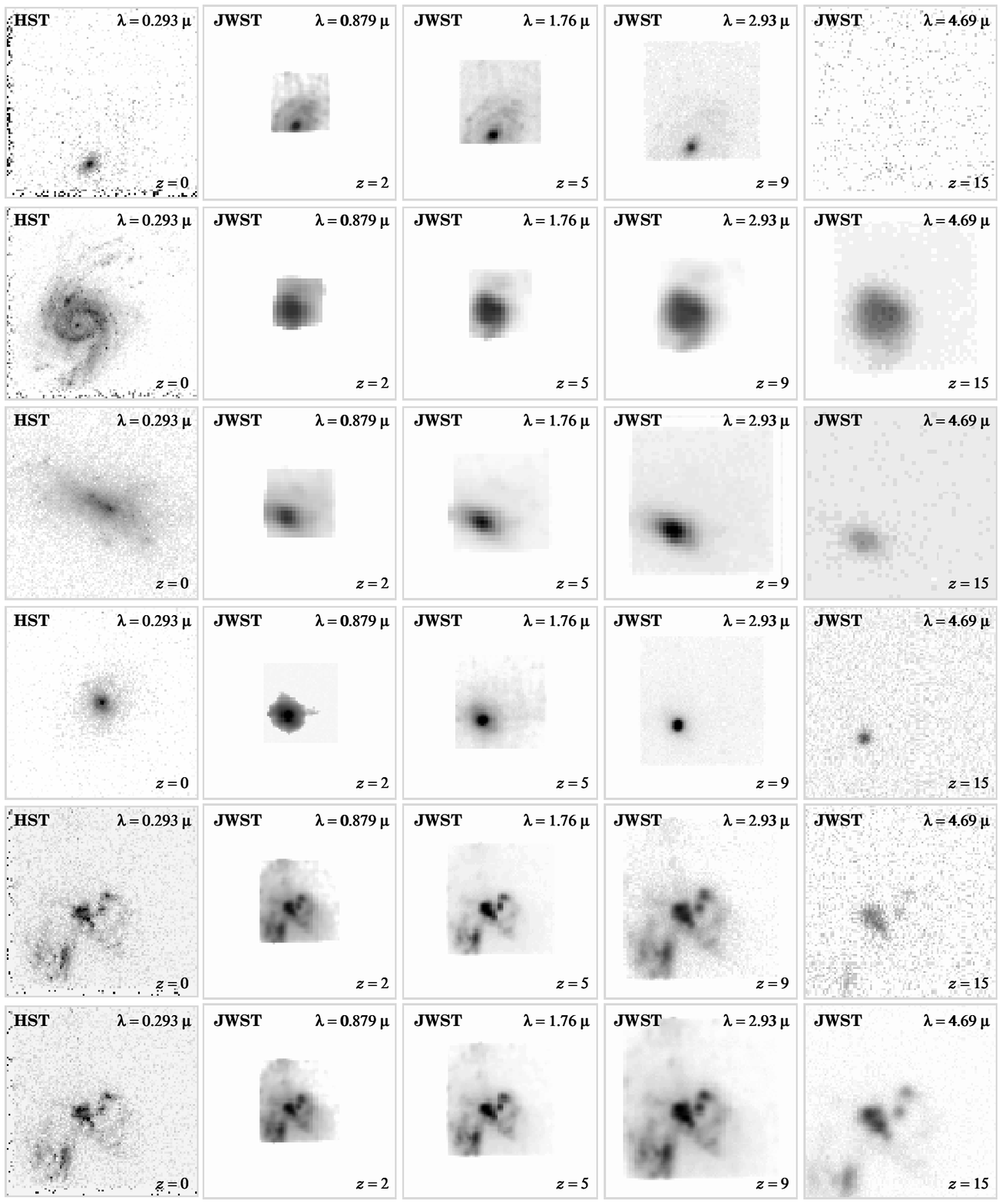,width=\textwidth}
}

\n {\footnotesize\baselineskip=10pt
{\bf Fig.~6.}\ Rows 1--5 show 1-hr JWST NIRCam simulations of nearby
galaxies observed with HST/WFPC2 in the mid-UV (Windhorst \etal 2002). 
Row 6 is a 100-hr simulation of JWST fully built to its specifications. 
The HST images (left column) are $37''$ or $75''$ across, and the JWST
images 0\arcspt 7--3\arcspt0.  The rows show that: (Row 1) Most disks
will SB-dim away at high z (but most disks formed at z$\,\simeq\,$1--2
anyway); (Rows 2--3 \& 5--6) High SB structures in star-forming objects
and mergers/train-wrecks are visible to very high z; (Row 4) Point 
sources (weak AGN) are visible to very high z.  The two high-SB clumps
in the upper right of the merger in Rows 5--6 have have 
$\sim$$10^8$--$10^9\Msun$ in stars, and are more representative of the
objects that JWST can detect at z\cge 10.  Deep JWST surveys can
quantitatively measure the evolution of galaxy structure and morphology
over the entire epoch of galaxy assembly.}
%%%%%%%%%%%%%%%%%%%%%%%%%%%%%%%%%%%%%%%%%%%%%%%%%%%%%%%%%%%%%%%%%%%%%%%%

\ve 

\n of cosmic time, by measuring their redshift distribution as a
function of rest-frame type (Driver \etal 1998; Fig.~5).  For this to
work, the galaxy structure must be well imaged for large samples from
deep, uniform and high quality multi-wavelength images, which JWST can
do through dedicated surveys. 

Spatially resolved NIRSpec and MIRI integral-field spectra of distant
galaxies --- when compared to the quantitative structure from NIRCam
Fourier Decompositions --- will allow to directly trace the physical
causes of locally enhanced star-formation: infall, bulk velocities in
excess of regular rotation, etc, and so map galaxy assembly in detail. 
For the sake of space, the detailed spatially-resolved spectroscopic
studies of distant galaxies possible with JWST NIRSpec and MIRI are not
reviewed here, but readers are referred to, e.g., Lilly \etal (1998) and
Abraham \etal (1999) as to how such studies have been done in the past,
and to the websites in the References as to how JWST will address the
mass assembly of galaxies in this manner more quantitatively, and at
much higher redshifts. 

Fig.~6 shows how --- with proper restframe-UV templates and training ---
JWST can quantitatively measure the evolution of galaxy morphology and
structure over a wide range of cosmic time.  These JWST simulations show
that: (1) Most disks become invisible due to SB-dimming at very high
redshifts (z$\,\simeq\,$15--20), but they likely formed at 
z$\,\simeq\,$1--2 anyway; (2) High-SB structures in star-forming objects
and mergers/train-wrecks are visible to z$\,\simeq\,$10--15; (3) Point
sources (e.g., Pop~III star clusters and weak AGN) are visible to 
z$\,\simeq\,$15--20.  With proper restframe-UV training, deep JWST
surveys can thus quantitatively measure the evolution of galaxy 
structure and morphology over the entire epoch of galaxy assembly (i.e.,
from z$\,\simeq\,$15 to z$\,\simeq\,$1). 

These simulations do not imply that observing star-forming objects at
z\cge 10 with JWST will be easy.  On the contrary, since galaxies formed
through hierarchical merging, many objects at z$\,\simeq\,$10--15 will
be 10$^1$--10$^4$$\times$ less luminous than the actively star-forming
objects seen with HST nearby, and hence require pushing JWST to its very
limits.  Beyond z\cge 10--12, the cosmological SB-dimming will also
render most of the lower-SB flux in objects invisible, except in the
longest JWST integrations, as shown in row~6 of Fig.~6.

\section{Conclusions and Afterword}

In summary, to have a decent chance of measuring the LF of the first
star-forming objects at z$\,\simeq\,$10--20 will require very long
integrations to AB=31.5~mag over a GOODS-sized field with the 6.5~m JWST
(Fig.~2b).  It is therefore important in the current definition and
re-planning of JWST, to keep its primary goals of detecting First Light
in mind. 

It is in this context prudent to briefly reflect on the history of JWST. 
Shortly after the first stunning refurbished HST results appeared in
1994, AURA prompted the community to outline what the future of space
imaging should look like.  This resulted in the ``HST and Beyond'' study
(Dressler \etal 1996) supported by NASA.  This committee recommended a
Large (4~m) Infrared-Optimized Space Telescope, then referred to as the
Next Generation Space Telescope (NGST).  The NASA Administrator Dan
Goldin subsequently told the community to think big, and consider a
6--8~meter deployable NGST.  In the mid-late 1990's, the earlier Adhoc
and Interim NGST Science Working Groups then outlined an initial science
program and requirements for the 8~m NGST (Stockman \etal 1997). 
Meanwhile, theory had sufficiently progressed to argue that the epoch of
First Light may have occurred at very high redshifts, and part of the
science case in the NGST Design Reference Mission was developed to
reflect this.  The 8~m NGST concept was endorsed with top priority by
the Decadal National Academy of Sciences survey (McKee \& Taylor 2001). 

In 2002, TRW (now Northrup-Grumman System or NGS) had the winning
proposal with a 7.0~m NGST, which was awarded the NASA contract.  Then
realism set in, and in late 2002, NASA and the Flight Science Working
Group (SWG) worked together to define the currently designed 6.5~m
telescope, named the James Webb Space Telescope in 2002.  In early 2003,
the WMAP polarization results suggested that First Light may have
occurred through Pop~III stars as early as z$\,\simeq\,$20.  While the
error bars on this result still have to come down and its redshift range
refined (N.~Wright, this Vol.), the Fe-lines seen in z$\,\simeq\,$6
quasars by Freudling \etal (2003) also suggested a very early epoch of
star-formation and supernovae.  In the last few years, the Spitzer Space
Telescope has shown us beautiful results underscoring the critical role
of dust in galaxies and optically hidden star-formation at high
redshifts, as shown in the special 2004 ApJ Suppl.\ Spitzer issue (e.g.,
Werner \etal 2004). 

In conclusion, if the JWST is to remain NASA's First Light machine, it
needs to have a 6.5~m class aperture and its near-IR to mid-IR
capabilities to assure proper measurement of the expected First Light
objects.  In a sense, JWST needs to combine the best of HST and Spitzer
and fold this into a mission that can successfully open the next major
frontier in astronomy. 

\n\small The HST work was supported by grants GO-8645.* and GO-9780.*
from STScI, which is operated by AURA for NASA under contract NAS
5-26555.  The JWST work was supported from NASA JWST grant NAG 5-12460. 
RAW thanks the other members of the JWST Flight Science Working Group,
the JWST Instrument Teams, and the JWST hardware teams for their
continuous dedicated work on the JWST project.  A PDF file of the review
presented at the conference is available at
\textsl{www.physics.uci.edu/Cosmology/\#schedule}\/.  This paper and
other JWST studies are also available as PDF files at: 
\textsl{www.asu.edu/clas/hst/www/jwst/}\/. \normalsize

\n

\end{document}